\documentclass[letter,twocolumn]{jpsj2} 

\def\ds{\displaystyle}
\def\bm#1{{\mbox{\boldmath $#1$}}}

\def\bms#1{{\mbox{\rm\scriptsize\it\bf#1}}}

\def\a{\alpha}

\def\g{\gamma}
\def\d{\delta}
\def\e{\varepsilon}

\def\m{\mu}

\def\r{\rho}
\def\s{\sigma}
\def\t{\tau}

\def\w{\omega}

\def\nn{\nonumber}

\def\del{\partial}

\title{
Calculation of Optical Conductivity, Resistivity and Thermopower of Filled Skutterudite CeRu$_4$Sb$_{12}$ based on a Realistic Tight-binding Model with Strong Correlation
}

\author{%
Tetsuro {\sc Saso}
}

\inst{%
Department of Physics, Saitama University, Saitama, 338-8570
}

\recdate{\hspace{3cm}}

\abst{
The filled-skutterudite compound CeRu$_4$Sb$_{12}$ shows a pseudo-gap structure in the optical conductivity spectra similar to the Kondo insulators, but metallic behavior below 80 K.  
The resistivity shows a large peak at 80 K, and the Seebeck coefficient is positive and also shows a large peak at nearly the same temperature.
In order to explain all these features, a simplified tight-binding model, which captures the essential features of the band calculation, is proposed.
Using this model and introducing the correlation effect within the framework of the dynamical mean field approximation and the iterative perturbation theory, the temperature dependences of the optical conductivity, resistivity and the Seebeck coefficient are calculated, which can explain the experiments.
}

\kword{filled-skutterudite, CeRu$_4$Sb$_{12}$, Optical Conductivity, Resistivity, Thermopower
}
\begin{document}
\maketitle
The filled-skutterudite compounds RT$_4$X$_{12}$ (R=rare earth, T=Fe, Ru, Os, X=P, As, Sb) show a variety of interesting properties, but their physical mechanisms are not well understood.  Among them, however, Ce-skutterudites show a clear systematic trend that X=P and As compounds are insulators whereas X=Sb yield metals.\cite{Sugawara05}  This is due to the larger atomic size from P to Sb, and hence larger hybridization with f electron on Ce in the X$_{12}$ cage.  Similar but weaker trend is seen also for T=Fe to Os.  In fact, all X=P compounds are insulators with a large energy gaps of the order of 0.1 to 0.03 eV, whereas CeRu$_4$Sb$_{12}$ is a metal at low temperatures according to the transport measurement\cite{Bauer01,Abe02}.  
Applied pressure renders CeRu$_4$Sb$_{12}$ to an insulator for $P>6$GPa, supporting the above picture.\cite{Kurita04}  

It is strange that nevertheless CeOs$_4$Sb$_{12}$ shows an insulating behavior at low temperatures, although the band calculation shows that it is a semimetal\cite{Harima03}.  Both Ru and Os compounds show similar pseudo-gap structures and mid-infrared peak in the optical conductivity spectra\cite{Dordevic01,Matsunami04} similar to the Kondo insulators,\cite{Okamura05,Saso04} but CeRu$_4$Sb$_{12}$ shows a metallic behavior below 80 K.  Therefore, It is an interesting issue whethere Ce-skutterudites with insulating behaviors at low temperatures are classified into the Kondo insulators or not, and what is a mechanism for the gap.  It was recently found that the insulating behavior of CeOs$_4$Sb$_{12}$ may be due to a formation of some ordered state (possibly spin density wave (SDW))\cite{Sugawara05}, to which we have proposed a theoretical model and calculation explaining the occurence of SDW and the anomalous phase diagram in this compounds.\cite{Imai05}   Therefore, we focus on CeRu$_4$Sb$_{12}$ in the present paper.  

We previously proposed a tigh-binding band model with full f-electron characters on Ce with the spin-orbit (s.o.) interaction and the crystal electric field (CEF), and a simplified A$_u$ top-most valence band composed of the p states on X$_{12}$ clusters.\cite{Mutou04} The calculated optical conductivity successfully explained the temperature dependence of the mid-infrared (IR) peak and the filling-up of the gap with raising temperature by inclusion of the strong correlation effects.  However, we have assumed a finite band gap in the previous work to focus on these temperature dependence of the (pseudo-)gap structures.  Also the band calculation gave a finite gap.\cite{Harima03}  It is now clear that CeRu$_4$Sb$_{12}$ is a metal at the ground state, perhaps a semimetal\cite{Abe02}, so that we improve our model with a slight simplification and calculate the transport properties in the present paper.

In CeRu$_4$Sb$_{12}$, the resistivity shows a large peak at 80 K after subtraction of the data of the La-counterpart.  Seebeck coefficient is positive and also shows a large peak at nearly the same temperature.\cite{Takeda00,Bauer01,Abe02}  Non-Fermi-liquid(NFL)-like behaviors are found at very low temperatures below several Kelvin,\cite{Bauer01,Takeda00} but we do not take them into our scope here.

In order to explain all these features except NFL, we propose a simplified tight-binding model, which captures the essential features of the band calculation.
Using this model and introducing the correlation effect within the framework of the dynamical mean field approximation and the iterative perturbation theory,\cite{Kajueter96} the temperature dependences of the optical conductivity, resistivity and the Seebeck coefficient are calculated, which can explain the experiments rather well.  Theoretical understanding of the transport properties of the skutterudite compounds is very important since they have attracted much interest because of its high potentiality for an application as a new good thermoelectric device.\cite{Sales96,Sales97}

The basic crystal structure of filled skutterudites is the body-centered cubic (bcc) lattice composed of the X$_{12}$ clusters filled by R in the center.  The f orbitals of rare-earth atoms strongly hybridize with the p orbitals on X$_{12}$ clusters close the Fermi level.  
The X$_{12}$ cluster has the T$_{\rm h}$ symmetry, but its effect on $J=5/2$ f$^1$ state in Ce is the same as that of O$_{\rm h}$.  Therefore, we use the notation for O$_{\rm h}$ in the followings.

The present band model consists of the single wide band (called c-band hereafter):
\begin{eqnarray}
\e_\bms{k}^c &=& t_1\cos\frac{k_xa}{2}\cos\frac{k_ya}{2}\cos\frac{k_za}{2} \nn \\
             & & +t_2(\cos k_xa+\cos k_ya+\cos k_za)
\end{eqnarray}
with the nearest and the next-nearest neighbor hoppings on the bcc lattice, which represents the top-most p band of X$_{12}$ clusters.
The lattice constenat $a$ is set equal to unity in the following.
For f-electrons, the neutron scattering experiment cannot resolve the CEF for the moment.\cite{Adroja03}  We here assume that the the ground state of f electron is $\Gamma_7$ doublet as in CeOs$_4$Sb$_{12}$,\cite{Bauer01b} and take the doubly degenerate f-band, but since CEF is not well separated, it might be the case  that the 6-fold J=5/2 states could be a better starting model.  We neglect this problem and take the simplest model, but the degeneracy may change the following results only quantitatively.
Then, we assume the following simple f-band: $\e_\bms{k}^f =E_{f} + \a \e_\bms{k}^c$, where $\a$ is a small factor representing the width of the f-band dispersion. This would be a reasonable choice since both X$_{12}$ and R occupy the same bcc sites.
These two c and f bands hybridize with each other.

Then Hamiltonian is given by
\begin{eqnarray}
H&=&\sum_{{\mathbf k}\sigma}\epsilon^{c}_{\mathbf k}
c^{\dag}_{{\mathbf k}\sigma}c_{{\mathbf k}\sigma}
+\sum_{{\mathbf k}\sigma}\epsilon^{f}_{\mathbf k}
f^{\dag}_{{\mathbf k}\sigma}f_{{\mathbf k}\sigma}\nonumber \\
&+&\sum_{{\mathbf k}\sigma}
\bigg(Vc^{\dag}_{{\mathbf k}\sigma}
f_{{\mathbf k}\sigma}+h.c\bigg)
+U\sum_{i}n^{f}_{i\uparrow}n^{f}_{i\downarrow}
\label{ham}
\end{eqnarray}
where $c^{\dag}_{{\mathbf k}\sigma}$($f^{\dag}_{{\mathbf k}\sigma}$) creates  a c (f) electron with the momentum ${\mathbf k}$ and spin $\sigma$ (representing $\Gamma_7$ doublet), and $n^{f}_{i\sigma}$=$f^{\dag}_{i\sigma}f_{i\sigma}$ with the site index $i$.
$V$ is the hybridization between conduction and f electrons.  Its ${\mathbf k}$-dependence is neglected.  These choices in our model band immensely facilitate the calculations of the strong correlation effects, which will be seen later.  Actually, of course, the f-dispersion is created through the X$_{12}$ clusters, so that it must be self-consistently determined with $\epsilon^{c}_{\mathbf k}$ and the hybridization $V_{\mathbf k}$.
$U$ is the Coulomb repulsion between f electrons.  

Without the Coulomb interaction, the diagonalized energy dispersions become $E_{\mathbf k}^{\pm}=[\epsilon^{c}_{\mathbf k}+\epsilon^{f}_{\mathbf k}\pm \{(\epsilon^{c}_{\mathbf k}-\epsilon^{f}_{\mathbf k})^{2}+4V^{2}\}^{1/2}]/2$, which are the so-called hybridized bands, but since $\epsilon^{f}_{\mathbf k}$ has a finite dispersion, the two bands can have finite overlap if $\alpha$ is not too small. 

\begin{figure}[tb]
\begin{center}
\includegraphics[width=7cm]{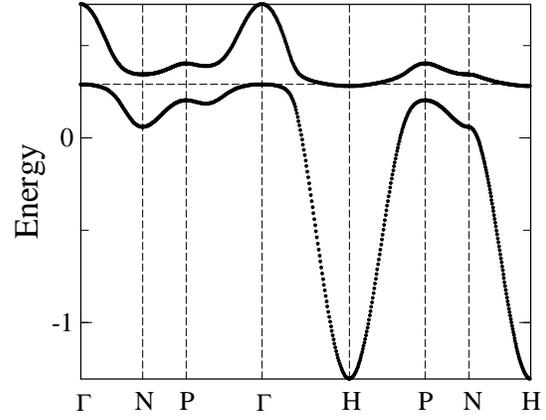}
\end{center}
\caption{
The model band structure of ${\rm CeRu_{4}Sb_{12}}$ calculated with the parameters $t_1$=1.0, $t_2=0.1$, $\alpha$=0.02, $E_{\rm f}$=0.3 and $V$=0.1 is shown. 
The dotted line represents the Fermi level. }
\label{Fig:band}
\end{figure}

\begin{figure}[tb]
\begin{center}
\includegraphics[width=7cm]{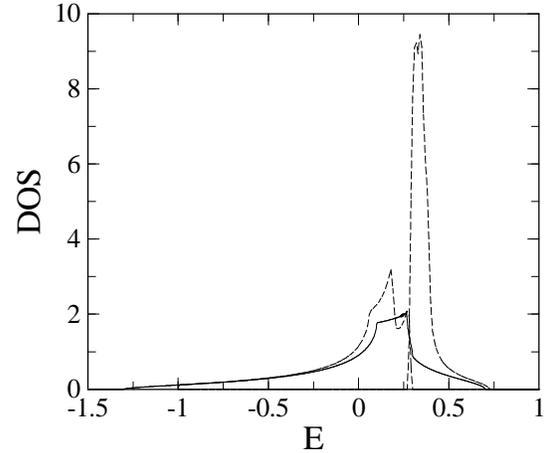}
\end{center}
\caption{
The density of states $\r_c(\e)$ for $\e^c_\bms{k}$ (solid line) and those for $E_\bms{k}^{\pm}$ (dotted lines) are shown. }
\label{Fig:DOS}
\end{figure}

Considering the empirical trend of the lattice constant and the energy gap of Ce-skutterudites, ${\rm CeRu_{4}Sb_{12}}$ must be a semimetal \cite{Sugawara05}. 
Our assumed band structure with $t_1$=1.0, $t_2=0.1$, $\alpha$=0.02, $E_{\rm f}$=0.3 and $V$=0.1 is shown in Fig. \ref{Fig:band}, which has a semi-metallic character: the top of lower band ($\Gamma$ point) and the bottom of upper band (H point) slightly touch the Fermi level, respectively, and overlap by about 0.03. 
Although our band structure is quite simplified, Fig. \ref{Fig:band} captures the low-energy structure of the band calculation for CeRu$_4$Sb$_{12}$\cite{Harima03p} rather well. The density of states (DOS) $\r_c(\e)$ for $\e^c_\bms{k}$ and those for $E_\bms{k}^{\pm}$ are shown in Fig.\ref{Fig:DOS}.

For treating the correlation effects, we use the dynamical mean-field theory and the iterative perturbation theory,\cite{Kajueter96} which is a second-order perturbation theory modified so as to interpolate between the weak and strong correlation limits.  Because of our choice $\e_\bms{k}^f =E_{f} + \a \e_\bms{k}^c$ and the $\bm{k}$-independent $V$, the $\bm{k}$-summation is converted into a single energy integration on $\e_\bms{k}^c$ with the density of states for this band, $\r_c(\e)$.  Therefore, the $\bm{k}$-summation over the Brillouine zone is necessary only once at the begining and for the calculation of the dynamical conductivity given by\cite{Mutou01}
\begin{eqnarray}
\label{eq:direct}
{\rm Re} \sigma(\omega) &=& 2\pi e^2 \bar{v}^2_{cx} \frac{1}{N} \sum_{\bms{k}} \int {\rm d}\e \frac{f(\e)-f(\e+\w)}{\w} \nn \\
 & \times &  \left[ \rho^c_\bms{k}(\e)\rho^c_\bms{k}(\e+\w) 
     + \a \rho^{cf}_\bms{k}(\e)\rho^{cf}_\bms{k}(\e+\w)\right. \nn \\
 & & \left. + \a^2 \rho^f_\bms{k}(\e)\rho^f_\bms{k}(\e+\w) \right].
\end{eqnarray}
Here we neglected the vertex correction and replace the velocity matrix elements by an average $\bar{v}_{cx}^2$ except its scaling factor $\a$, 
$f(\e)$ the Fermi function and $\rho^\g_\bms{k}(\e)=(-1/\pi){\rm Im}G^\g_\bms{k}(\e+i\d)$ ($\g=$c or f or cf) with the Green functions
\begin{eqnarray}
\label{eqn:k-Green}
G^c_{\bms{k}}(\e) &=& \frac{1}{\e  - \e^c_{\bms{k}} - \frac{\ds V^2}{\ds \e-\e^f_{\bms{k}}-\Sigma_f(\e)}}, \\
G^{cf}_{\bms{k}}(\e) &=& \frac{V}{(\e  - \e^c_{\bms{k}})(\e-\e^f_{\bms{k}}-\Sigma_f(\e)) -  V^2}, \\
G^f_{\bms{k}}(\e) &=& \frac{1}{\e  - \e^f_{\bms{k}} -\Sigma_f(\e)- \frac{\ds V^2}{\ds \e-\e^c_{\bms{k}}}}.
\end{eqnarray}

When the imaginary parts of the self-energy and $\a$ are small, Re$\s(\w)$,  are dominated by the current carried by the c-electrons and can be approximated by\cite{Stojkovic97}
\begin{eqnarray}
  \mbox{Re}\s(\w) &\simeq&  2\pi e^2 \bar{v}^2 \sum_\bms{k} \frac{f(\e^c_\bms{k})-f(\e^c_\bms{k}+\w)}{\w} \nn \\
  & & \times {\rm Im} \frac{1}{\w+\Sigma_c^R(\e^c_\bms{k},\bm{k})-\Sigma_c^A(\e^c_\bms{k},\bm{k})},
  \label{eq:opcon-a}
\end{eqnarray}
which is not a bad approximation.
Here,
\begin{equation}
  \Sigma_c(\e,\bm{k}) \equiv \frac{V^2}{\e-\e^f_\bms{k}-\Sigma_f(\e)},
\end{equation}
and R (A) denotes the retarded (advanced) function.

The resistivity $\r(T)$ and the Seebeck coefficient $S(T)$\cite{Jonson90} for correlated systems are given by
\begin{eqnarray}
  1/\r(T) &=& 2e^2 \int d \e L(\e) \left( -\frac{\del f}{\del \e}\right), \label{eq:Res} \\
  S(T) &=& - \frac{1}{eT}\int d \e L(\e) (\e-\m) \left( -\frac{\del f}{\del \e}\right) \nn \\
   & & \hspace{5mm} / \int d \e L(\e) \left( -\frac{\del f}{\del \e}\right), \label{eq:TEP}
\end{eqnarray}
where $L(\e) \equiv (\pi N)^{-1}\sum_\bms{k} (v_{\bms{k}x})^2 \left[\mbox{Im}G^c_\bms{k}(\e,\bm{k})\right]^2$ when the c-electron dominates and the vertex corrections can be neglected.  It can be further approximated by $L(\e) \simeq \r_c(\e) \bar{v}_{cx}^2 \t_c(\e)$ similar to eq.(\ref{eq:opcon-a}) with the relaxation time of c electrons: $\tau_c(\e^c_\bms{k})^{-1} \simeq -2\mbox{Im}\Sigma_c(\e^c_\bms{k},\bm{k})$.  Both approximations yield the similar results, as will be shown below.

\begin{figure}[tb]
\begin{center}
\includegraphics[width=7cm]{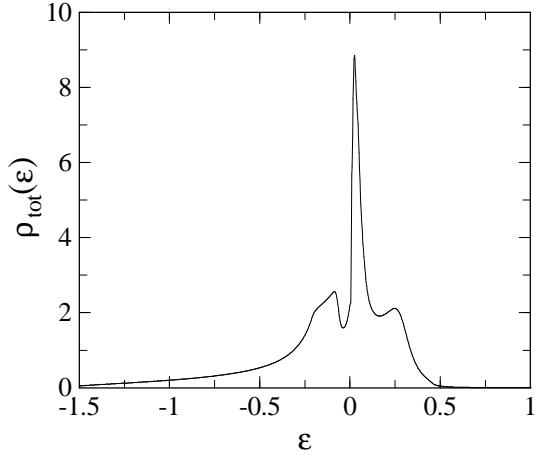}
\end{center}
\caption{
The total quasi-particle density of states for $U=0.3$ and $T=0$ is shown.  The origin of the energy is shifted to the Fermi level.}
\label{Fig:q-DOS}
\end{figure}

We numerically solve the f-electron self-energy $\Sigma_f(\e)$ for $U$=0.3 by iteration, and calculate the above-mentioned quantities. The $\bm{k}$-summations are carried over $400^3$ mesh points in the 1/8 of the extended cubic Brillouin zone.
Because of the correlation effects, the quasi-particle DOS becomes temperature-dependent, and the sharp Kondo peak grows up on the Fermi level at low temperatures in addition to the upper Hubbard peak (Fig.\ref{Fig:q-DOS}).  The appearance of the Kondo peak enhances the temperature-dependence of the physical quantities compared to the rigid band calculation.

\begin{figure}[tb]
\begin{center}
\includegraphics[width=8cm]{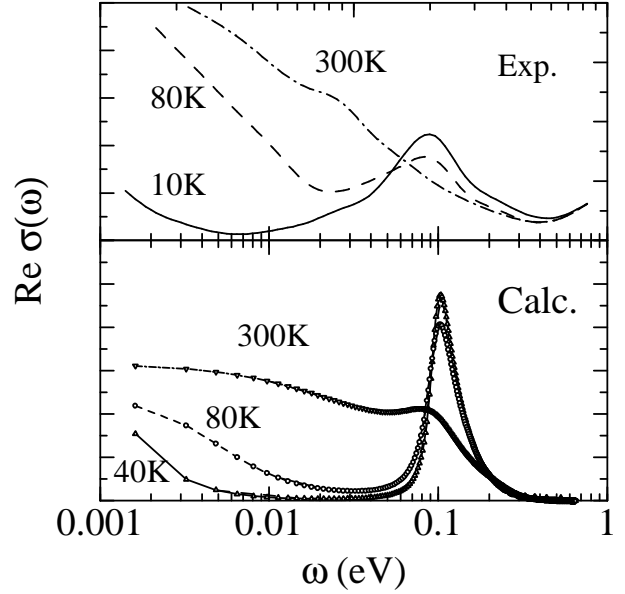}
\end{center}
\caption{
The experimental\cite{Dordevic01} and theoretical results for the dynamical conductivity for various temperatures are shown.}
\label{Fig:opt}
\end{figure}

Choosing the scale as $t_1=7000$ K, the calculated optical conductivity (by the correct formula, eq.(\ref{eq:direct}), but with only the cc-part) in Fig.\ref{Fig:opt} shows a mid-IR peak at 0.1 eV in accord with the experiment, but the width is narrower.  There is a large pseudogap below the mid-IR peak which is due to the direct ($\bm{k}$-conserving) interband transition, but a Drude peak at low temperature and low frequencies is seen since it is a semimetal.  The main descripancy is that the experimental spectrum at the lowest temperature has a long tail below the mid-IR peak, which may be due to the indirect transitions, as was pointed out recently by Okamura, et al. and the present authorfor the Kondo insulator YbB$_{12}$.\cite{Okamura05,Saso04,Mutou04}
The temperature-dependence is weaker than the experiments, but the psuedo-gap of the order of 0.1 eV is filled up already at $T=300$K, which is due to the many-body effect.\cite{Saso04,Mutou04}  Despite some descripancies, the overall structures and the tepmerature-dependence are rather well reproduced by the present simple model.  

\begin{figure}[tb]
\begin{center}
\includegraphics[width=8cm]{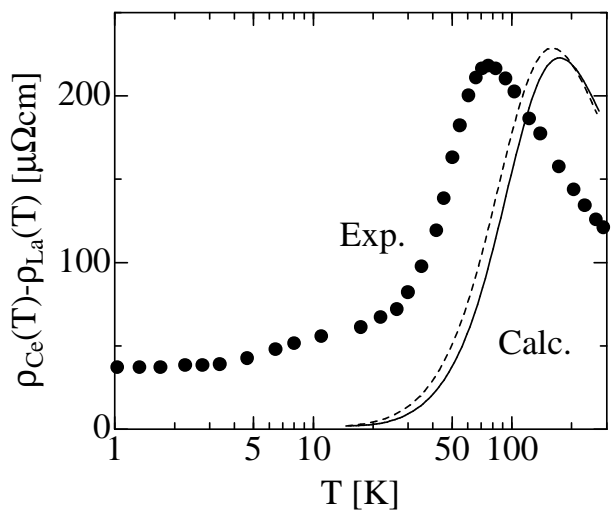}
\end{center}
\caption{The calculated resistivity (the broken line by the relaxation time approximation) is compared with the experimental value is shown. (The data of the coressponding La-compound is subtracted from the experimental data.\cite{Abe02})
 }
\label{Fig:res}
\end{figure}
\begin{figure}[tb]
\begin{center}
\includegraphics[width=8cm]{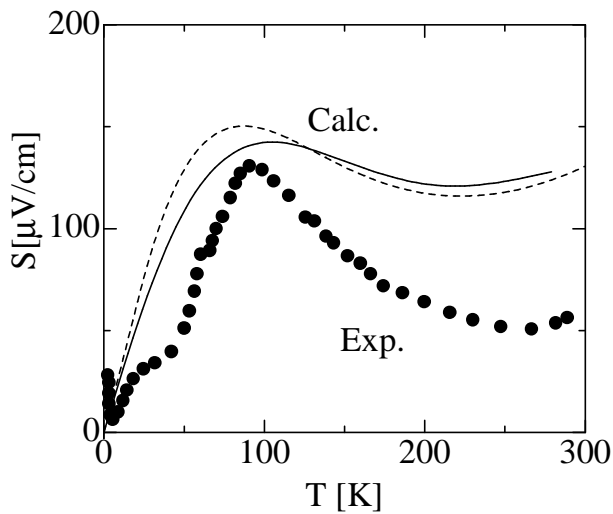}
\end{center}
\caption{The calculation for the Seebeck coefficient (the broken line by the relaxation time approximation) is compared with the experiment.\cite{Abe02}
 }
\label{Fig:Seebeck}
\end{figure}

The resistivity and the Seebeck coefficient with and without the relaxation approximation are compared with the experiment in Figs.\ref{Fig:res} and \ref{Fig:Seebeck}.  The calculated peak position agrees with the experiment in the Seebeck coefficient, but is twice higher in the resistivity. Nevertheless, our results capture the global features rather well, except the strange steep rise of $S(T)$ at very low temperatures and the shoulder at 20 K.  The difference in $\r(T)$ at low temperatures may be due to the residual resistivity.

We tried to fit all these data by the present model by changing the band parameters and $U$, but the present ones are the best.  The remaining discrepancies may be due to the oversimplification of the present band model, improvement of which is left for a future study.  

In summary, we have constructed the effective tight-binding model (periodic Anderson model with dispersive f band) to describe the low energy state of CeRu$_4$Sb$_{12}$, and calculated the optical conductivity, the resistivity and the Seebeck coefficient, taking account of the Coulomb interaction between f electrons via the dynamical mean-field approximation and the iterative perturbation theory.  The results can explain the experimental data semi-quantitatively.  Improvement of the model and the consideration of the non-Fermi-liquid-like behaviors found at low temperatures remain to be done in the future.

We would like to thank Professors H. Okamura, H. Sato and H. Harima for useful informations and discussions.  
This work was supported by The Grant-in-Aid from the Ministry of Education, Science and Culture: ``Evolution of New Quantum Phenomena Realized in the Filled Skutterudite Structure'', No.  16037204.

\end{document}